# Particle Physics Masterclasses

K. Cecire, QuarkNet National Staff Teacher
*University of Notre Dame, Notre Dame, IN 46556, USA*

The IPPOG and U.S. Particle Physics Masterclasses took place worldwide in March 2011. For the first time, all masterclasses used real LHC data. Students in the U.S. Masterclasses (that included participants in several countries outside the U.S.) analyzed both ATLAS and CMS data. QuarkNet has been evaluating the U.S. effort since 2008. The design of the LHC masterclasses and the results of this study will be discussed.

## 1. Introduction

Particle physics masterclasses are a chance for high school students to visit universities or labs to be "particle physicists for a day." These experiences are like masterclasses in the arts, in which students prepare works so that experts can mentor the students so they can improve these works and learn technique. In particle physics masterclasses, high school physics students learn about collider physics and then analyze real CERN data in the form of event displays. The mentors are particle physicists who help the students understand the data, the detector, and the meaning of any conclusion. A masterclass in music might end with a concert; a particle physics masterclass ends with a videoconference between masterclass institutes that have met that day in which students compare results and ask questions.

In Europe, there are usually four to six masterclass institutes, which meet on a given day in February or March of each year over a three- to four-week interval. The videoconferences are moderated by physicists at CERN and feature a light quiz to add interest.

In the U.S. program, two to four institutes meet per day over a one- to two-week interval and the moderation is done from Fermilab. The U.S. program does not include a quiz.

## 2. LHC Masterclasses

### 2.1. A New Probe (for physics)

The European Particle Physics Outreach Group (EPPOG, now IPPOG, where the I is for International) instituted International Particle Physics Masterclasses in 2005 using a paradigm created by physicists in the U.K. and having students study event displays of Z and W boson decays. Data was taken from the OPAL and DELPHI experiments in the Large Electron-Positron Collider (LEP). Until 1999, LEP occupied the tunnel, which now houses the Large Hadron Collider (LHC).

In March 2011, real data from the LHC was exclusively used in the International Masterclasses. There were data exercises for the ATLAS, ALICE, and CMS detectors.

### 2.2. ATLAS

Students can choose from two exercises in the ATLAS masterclass [1]. The first is the "W-path," in which they probe the structure of the proton by examining the W+/W- ratio from data. The other is the "Z-path," in which students examine a variety of events and filter out all but dilepton Z decays to create a mass plot. U.S. Masterclass Institutes with an ATLAS orientation focused on the Z-path.

The W-path exercise uses the MINERVA software developed at Rutherford-Appleton Laboratories in the U.K. while the Z-path uses HYPATIA software developed at the Universities of Athens and Belgrade. Both software packages are downloaded along with data to local computers prior to the masterclass and both are based on the Atlantis event viewer developed for ATLAS. Figure 1 shows a sample from HYPATIA.

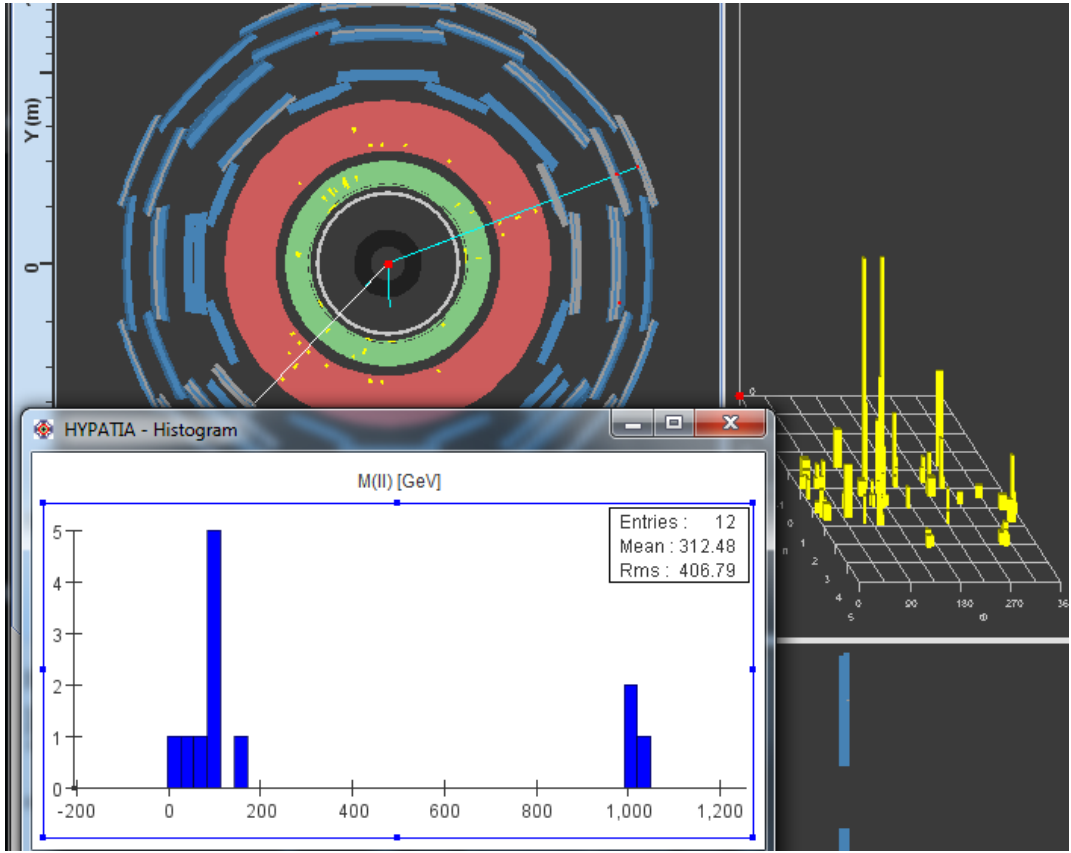

Figure 1: HYPATIA event display (background) for ATLAS Masterclass with a data histogram (foreground). Note one peak around the Z mass of 91.4 GeV from real ATLAS data; the other is around 1000 GeV for a hypothetical Z' from Monte Carlo data.

### 2.3. ALICE

The ALICE experiment developed a software package for the masterclasses based on the ROOT environment. Students develop mass plots of $\Lambda$ and $K_0^S$ particles in the search for strangeness in LHC data.

### 2.4. CMS

The CMS masterclass [2] used J/Ψ data available from CMS; this data was used along with an online, simplified version of the CMS iSpy event display to create a masterclass exercise. The event display, called "iSpy-online," has the advantages of working entirely in a browser via the Web with nothing to download and that it can be manipulated in three dimensions. In the CMS J/Ψ exercise, students study dimuon events with calculated invariant masses in the 2 GeV to 5 GeV range and determine the quality of the events based on the muon tracks; higher quality events are more likely to be J/Ψ particles than background. Thus the students filter the events in order to get the best possible mass plot. (Unfiltered data shows a J/Ψ mass peak, which is hard to distinguish from noise.) Figure 2 shows iSpy-online with a sample mass plot and masterclass moderators.

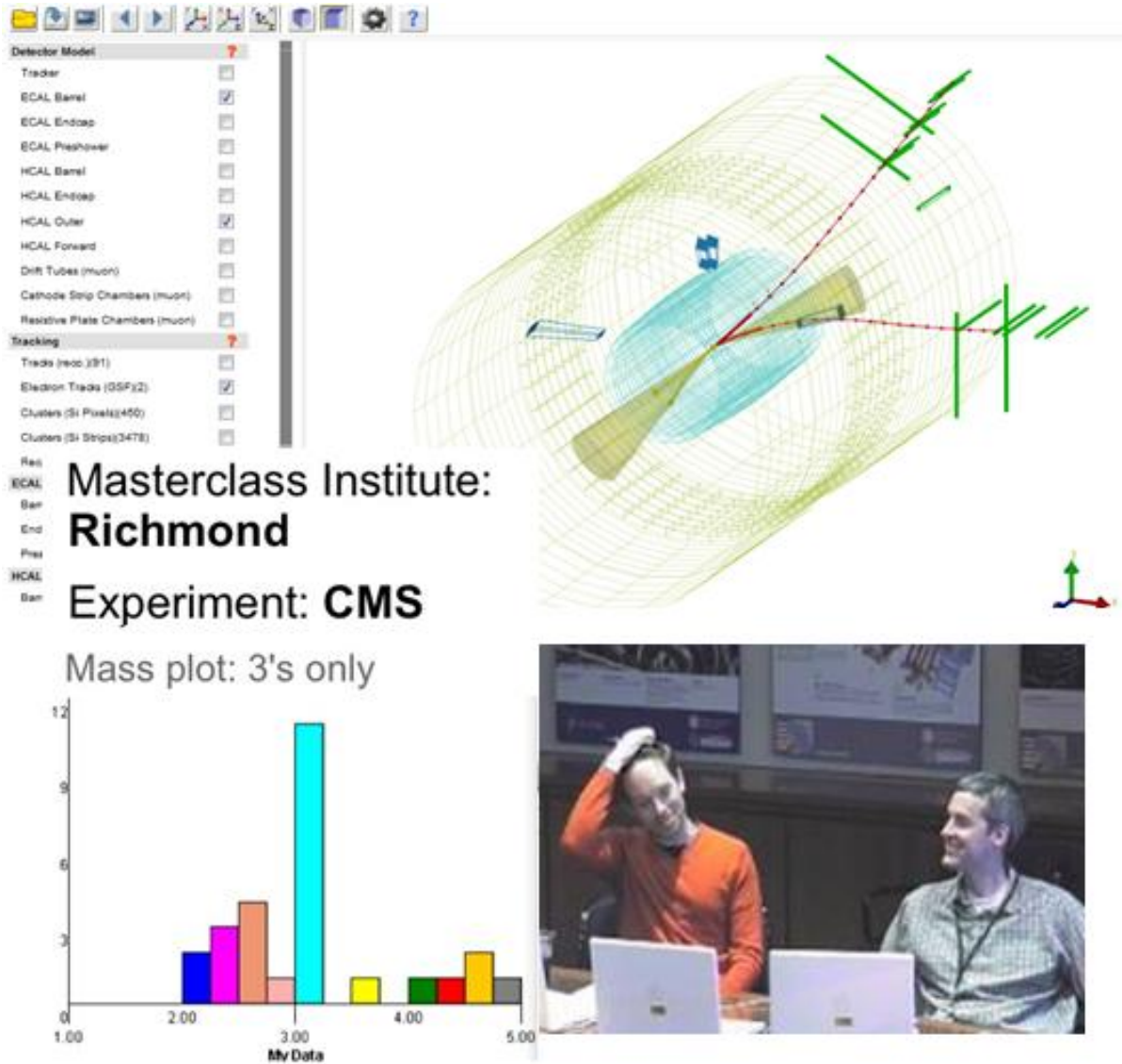

Figure 2: CMS iSpy-online event display showing a J/Ψ → μ⁺μ⁻ candidate event (top); mass plot of the J/Ψ from actual CMS data done in a masterclass after students have filtered the events (bottom left); masterclass moderators Michael Kirby (left) and Jake Anderson (right) in a videoconference at Fermilab.

## 3. Results and Future

### 3.1. Results from Evaluation and Observation

A survey of U.S. Masterclass students in 2011 compared with 2010 revealed that they were generally satisfied with the program. There was some improvement in this overall satisfaction between the two years; this is important and heartening in light of the change from LEP to LHC exercises.

Observations from mentors and teachers indicated that putting ATLAS and CMS masterclass institutes together in the same videoconference caused problems. The greatest of these was that students did not understand terminology and the exercise from the "other" experiment and were therefore confused; if the discussion in the videoconference was about ATLAS results, the students in CMS masterclass institutes could lose interest.

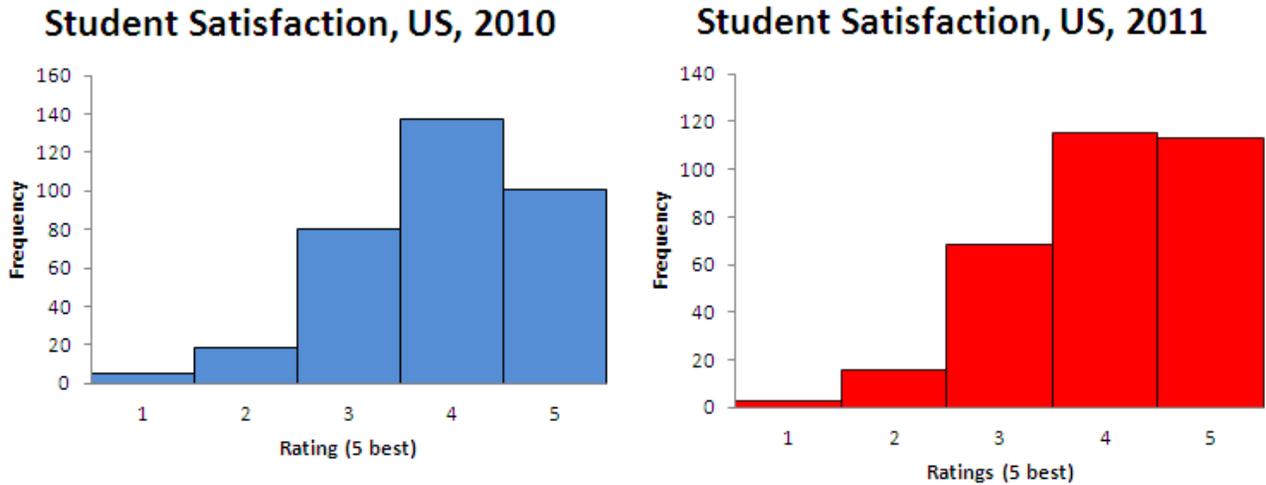

Figure 3. Student responses to the question, "How did you like the particle physics masterclass you attended today?" by U.S. students in 2010 and 2011. The rating scale was from 1 (Not at All) to 5 (A Lot).

### 3.2. Future Prospects and Plans

Moving forward into the 2012 masterclasses, the main thrust of using LHC data will not change. However, there are significant developments nonetheless. First, there is now more data available for the CMS masterclass and a new W/Z exercise is being developed. This will include students measuring key ratios ($W^+/W^-$ and W/Z) and creating Z mass plots from dilepton events. The events will, broadly, be for calculated invariant masses below 120 GeV and will all come from CMS data.

The ATLAS masterclass will expand as well, adding the use of data on lepton directions to probe W bosons more deeply.

### Acknowledgments

This work is supported in the United States through QuarkNet [3] and funded by the U.S. National Science Foundation and the U.S. Department of Energy, Office of Science. The author wishes to acknowledge the work of the group at the Technical University of Dresden, funded in part by the Helmholtz Alliance and the European Physical Society, in organizing the International Masterclasses under the auspices of the International Particle Physics Outreach Group. The author is grateful for the support and assistance of teachers, staff, and principal investigators in QuarkNet as well as colleagues at the University of Notre Dame.